\documentclass[english]{article}
\usepackage[T1]{fontenc}
\usepackage{graphics}
\usepackage{babel}
\usepackage{a4}
\makeatother
\providecommand{\LyX}{L\kern-.1667em\lower.25em\hbox{Y}\kern-.125emX\@}

\begin{document}
\title{\textbf{\large Large pion pole in \protect\(
Z_{S}^{MOM}/Z_{P}^{MOM}\protect \)
from Wilson action data}\large }

\author{J.R. Cudell\protect\( ^{a}\protect \), A. Le Yaouanc\protect\(
^{b}\protect \)
and C. Pittori\protect\( ^{c}\protect \)}

\maketitle
\begin{quote}

\( ^{a} \)Institut de Physique, Université de Liège, Sart Tilman, B-4000 Liège,
Belgique,\\ JR.Cudell@ulg.ac.be

\( ^{b} \)LPTHE, Univ. de Paris XI, Centre d'Orsay, F-91405 Orsay, France
\footnote{Unit\'e Mixte de Recherche - CNRS - UMR 8627},
\\ Alain.Le-Yaouanc@th.u-psud.fr

\( ^{c} \)Università di Roma II ``Tor Vergata'', Via della Ricerca Scientifica
1, I-00133 Rome, Italia, Carlotta.Pittori@roma2.infn.it
\end{quote}

\begin{abstract}
We show that, contrarily to recent claims, data from the Wilson (unimproved)
fermionic action at three
different \( \beta  \) values demonstrate
 the presence of a large Goldstone
boson contribution in the quark pseudoscalar vertex, quantitatively close to
our previous
estimate based on the SW action
with $c_{SW}=1.769$. We show that
discretisation errors
on $Z_{S}^{MOM}/Z_{P}^{MOM}$ seem to be much smaller than the Goldstone pole
contribution over a very large range of momenta. The subtraction of this non
perturbative contribution leads to numbers close to one-loop BPT.
\end{abstract}
\noindent LPT-Orsay 00-122 (november 2000) \par

\section{Introduction}

In a recent paper \cite{PS}, using data\footnote{
To calculate $Z_\psi$, we have also used the propagator data at \( \beta
=6.0
\)
kindly communicated by the Rome group.
} from the QCDSF collaboration
at \( \beta =6.0 \) and \( c_{SW}=1.769 \) for three different values of \(
\kappa  \)
\cite{QCDSF}, we have shown that the quark pseudoscalar vertex contains an
unexpectedly large contribution from the Goldstone boson pole,
{and that this contribution accounts for a third of $Z_P$
at
 $2~{\textrm {GeV}}$}. Qualitative indication
of this large Goldstone contribution had been noted previously by the QCDSF
group \cite{QCDSF-pole}, and by the Rome group \cite{ROME95}. A large
quantitative  estimate has been found independently
by JLQCD \cite{JLQCD} with staggered fermions.

Giusti and Vladikas 
\cite{giusti1}
have recently presented a criticism of our paper; they have reexamined this
problem using
Wilson data at two values of $ \beta= 6.2,\ 6.4$,
and have come to the conclusion
that the Goldstone boson term would be below the level of discretisation errors
"around \( p=1/a \)", and therefore not significant. However, they do not
compare their data with our result $-$ given at $2~{\textrm {GeV}}$
($a p \simeq 1$ at $\beta=6.0$)$-$
 but rather consider higher momenta $p=3.3 ~{\textrm {or}}~4.6~{\textrm {GeV}}$
($sin^2 (ap)=0.8$ in their Fig. 4 and Table 1), where of course the Goldstone
is
much smaller\footnote{One should not speak of the magnitude of the Goldstone
at \( p=1/a \) independently of the value of $\beta$, since this magnitude
depends on $p$, and $p=1/a$ depends of course on $\beta$.
{We have never spoken ourselves} of such a thing, as
stated in the abstract of \cite{giusti1}, but of its magnitude at $2~{\textrm
{GeV}}$ ($\beta=6.0$).}. We present here an analysis
based on a set of previously published data of the same origin \cite{giusti2},
which shows that, contrary to their objections, the Goldstone boson
contribution in Wilson data is in fact
completely compatible with our previous estimate, and much above discretisation
errors at $2~{\textrm {GeV}}$ and probably at notably higher momenta\footnote{
The present study could presumably be improved using the raw data of
\cite{giusti1},
which were unfortunately not available to us up to now. We now hope to
{improve our results} in the future with the help of the
authors of \cite{giusti1}.
}.

The present study in fact improves our determination of the Goldstone pole.
First of all, the set of  Wilson data analysed here is obtained at
larger values of \( \beta  \), up to \( \beta =6.4 \), where discretisation
errors should become really small at moderate \( p\simeq 2~{\textrm{GeV}} \)
(\( a^{2}p^{2}\simeq 0.25 \) at $\beta=6.4$), and perhaps better than with the
previous \( \beta =6.0 \)
data \cite{QCDSF} with ALPHA SW improved action. Secondly, we are now in a
position to
give a reliable estimate of the discretisation errors
by considering {the evolution of the
parameters with \( \beta  \)}.
{Thirdly, thanks to the $\beta=6.2, 6.4$ data,} we can improve the large
momentum tail of our
analysis of power corrections. Finally, we shall also improve
our previous work by the inclusion of statistical errors.

Having shown that the Wilson data confirm rather beautifully our first
estimate, exhibiting
\textit{\emph{a remarkable stability of the effect with increasing}} \textit{\(
\beta  \)},
we shall conclude by a critical analysis of the procedures of \cite{giusti1}
which lead to erroneous conclusions.

\section{Previous results on the Goldstone pole in the pseudoscalar vertex}

\subsection{Our previous results}

The theoretical expectation from the continuum is that a pole in \( 1/m_{q} \)
must be present in the pseudoscalar quark
vertex at $q=0$, as a consequence of the existence of the Goldstone boson. In
\cite{PS}, we analysed the lattice data kindly communicated by QCDSF
collaboration \cite{QCDSF} for the PS vertex at \( \beta =6.0 \) with SW
action at $c_{SW}=1.769$, at several \( \kappa  \), combined with propagator
data from the Rome
group, at the same \( \beta  \)
with the same action. We have obtained,
in the MOM renormalisation scheme,  a Goldstone-like fit of 
$\left(Z_{P}^{-1}\right)_{MOM}
=\Gamma _P/Z_\psi$
as function of \( \kappa  \). Namely, we have
shown that, at \( ap=1 \) and $\beta=6.0$, i.e.
around \( 1.9-2 \) GeV:
\begin{eqnarray}
Z_{P}^{-1}(2~{\textrm{GeV}})=1.88+\frac{0.023}{am_{q}}. & \label{notreZP}
\end{eqnarray}
 The first term on the r.h.s. is the ($\beta$-dependent) short-distance
contribution. The Goldstone pole corresponds to
the second term\footnote{
Recently, on investigating the quark propagator and the Ward identity relating
the PS and the propagator \cite{propagator}, we have improved the precision of
the determination of $Z_P$,and obtained similar numbers.
However, we shall stick here to our first determination, to which
\cite{giusti1} is referring.}, i.e. a pole in \( m_{q} \) at \( m_{q}=0 \). We
have also
checked that, as function of \( p^{2} \), one has the expected behavior: the
short distance term is compatible with a logarithmic dependence
\( \sim \left[\alpha_{s}(p^2)\right]^{4/11} \)
and the Goldstone term has a \( 1/p^{2} \) decrease.
Converting to physical
units, with \( a^{-1}=1.9~{\textrm{GeV}} \), {one obtains:}
\begin{eqnarray}
Z_{P}^{-1}(p^2)=Z_{P}^{-1}(short\; distance)+{0.158\: {\textrm{GeV}}^{3}\over
m_{q}\: p^{2}} & \label{notreZP2}
\end{eqnarray}
Of course, the effect of the uncertainty due to the error on \( a^{-1} \)
could be relevant for the Goldstone contribution since it is
\( \propto a^{-3} \). Despite this fact, and despite the presence of other
uncertainties, it seems difficult to escape the conclusion that the magnitude
of the Goldstone term is large at the smallest quark mass (around \( m_{q}=50
\)
MeV) and at \( 2 \) GeV: \( 30\% \) of the total \( Z_{P}^{-1}=2.7 \) at
\( 1.9 \) GeV, although \textit{\emph{it is decreasing rapidly with
increasing}}
\textit{\( p^{2} \)}. The result can be translated into an estimate of the
Georgi-Politzer mass at 1.9 GeV in the chiral limit: \( m_{R}=34 \)
MeV.

\subsection{Related findings of JLQCD and ALPHA}

Our evaluation (\ref{notreZP},\ref{notreZP2}) is quantitatively supported by
the remarkable JLQCD results on the pseudoscalar vertex and the mass operator
with staggered fermions \cite{JLQCD}. These results are important as they
benefit
from two advantages: they go down to very small quark masses (about 20 MeV),
and they have, in principle, small discretisation errors. The phenomenon
appears
very stable with respect to \( \beta  \) as they considered \( \beta =6.0,\:
6.2 \)
and 6.4.

The above estimate of the Goldstone term is also supported by the estimate of
the ALPHA group for the short distance \( Z_{P} \) \cite{ALPHA}, which must
be considered as very solid, since they work at ultra-short distances, and
since
their discretisation errors are very well controlled. When their short-distance
result is converted into the MOM scheme\footnote{The initial idea of this
conversion is due to Vittorio Lubicz.} and evolved perturbatively (at 3 loops
) down to \( 2 \) GeV, one obtains \( Z_{P}^{-1}(2\:
{\textrm{GeV}})=1.8 \).
This result is close to the first term of Eq.(\ref{notreZP}), and
\textit{\emph{quite
different from the total}} \( Z_{P}^{-1}(2\: {\textrm{GeV}})\simeq 2.5-2.7 \):
\textit{\emph{the difference must be filled by the Goldstone boson pole,
unless there be
 incredibly large discretisation errors in the total $Z_{P}^{-1}$}} \emph{.}
The latter
is very \emph{}unlikely in view of the following discussion of Wilson data.

\section{The Goldstone pole in Wilson data}

The most valuable part of  \cite{giusti1} is the introduction of the ratio \(
(Z_{P}/Z_{S})^{RI/MOM}=Z_{P}^{MOM}/Z_{S}^{MOM} \) and, on the other hand, of
some interesting Ward-Takahashi identities. Let us emphasize indeed that
$Z_{P}^{MOM}/Z_{S}^{MOM}$ is a scale-dependent quantity, in contrast
to \( (Z_{P}/Z_{S})^{WI} \), but with a \textit{\( p^{2} \)} dependence due
only
to power corrections \textit{\emph{}}.
{This gives it an important advantage} over \( Z_{P} \)
which necessarily contains a purely perturbative
contribution with logarithmic behaviour, complicating the determination
of the power term.

We therefore discuss \( Z_{P}^{MOM}/Z_{S}^{MOM} \)
as \textit{the best probe of the Goldstone pole}, which should be seen as a
\( {1/m_{q\: }p^{2}} \) term in the \textit{inverse}, \(
Z_{S}^{MOM}/Z_{P}^{MOM} \). Moreover, we shall show later that, according to
equation (19) of \cite{giusti1}, a \textit{\( p^{2} \)} change in \(
Z_{P}^{MOM}/Z_{S}^{MOM} \) can only be due to the presence of
a Goldstone \( 1/m_{q} \) pole.
The physical source of any departure from the $p^2\rightarrow\infty$ 
asymptotic value must then be
a Goldstone contribution. That it is present is recognized in \cite{giusti1};
we differ on the estimate of its magnitude.

The first question to be answered is whether the effect of the Goldstone
pole has the large magnitude that we have estimated, or whether it is
sub-dominant
{with respect to} discretisation errors already at \( p=2\)~GeV, 
as claimed in \cite{giusti1}.
This can be answered only by considering the behaviour of \(
Z_{S}^{MOM}/Z_{P}^{MOM} \)around
\( 2 \) GeV, or, in a scale independent manner, by comparing the
\textit{coefficient}
of the power corrections to the one we have given in Eq. (\ref{notreZP2}).

The second point concerns the estimate of the discretisation error itself: its
magnitude can
be estimated by {examining the
stability} of the result as a function
of \( \beta  \) with \emph{}\textit{\emph{fixed physical parameters}}\emph{.}

\subsection{Methodological considerations}

{\subsubsection{Use of physical units and comparison of different
actions}}

We have to compare different sets of data, with different actions and different
\( \beta  \)'s. It is a delicate task, especially for scale-dependent
quantities.
Since the Goldstone pole residue is a physical effect
{$-$ although perhaps gauge
dependent $-$} seen in the renormalised pseudoscalar (\( PS \)) vertex, a
minimum
requirement is to compare the results at
{\textit{identical momenta for the same quark
mass}}, not at identical \( ap \) if \( \beta  \) varies, as done
in ref. \cite{giusti1}, e.g. when making statements about the magnitude of the
Goldstone as compared to ours, at ``$p=1/a$". Our Fig. 1 below illustrates the
effect of comparing data in terms of $a p$
instead of $p$: the data at various $\beta$'s, which show large discrepancies
in terms of $a p$ (Fig. 1 a), almost superpose in terms of physical $p$ (Fig. 1
b)\footnote{This is also quite visible in Figs. 3 to 5 of
\cite{giusti2}.}.
\bigskip{}
\vspace{0.3cm}
{\par\centering \resizebox*{0.49\textwidth}{!}{\includegraphics{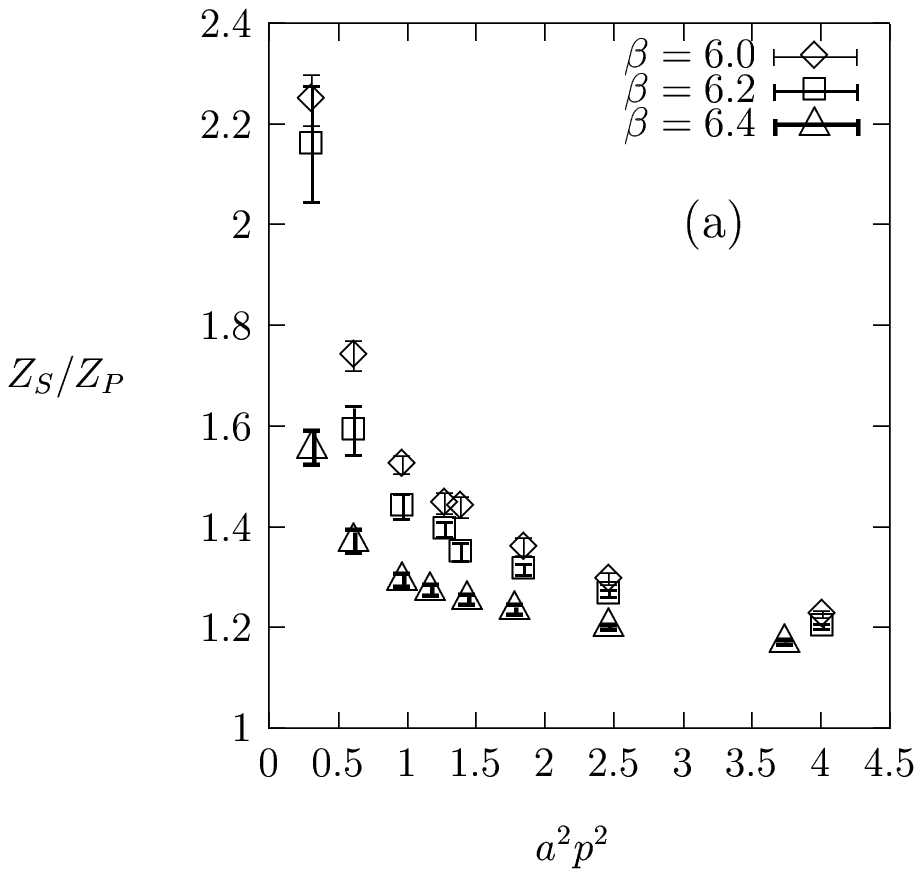}}
\resizebox*{0.49\textwidth}{!}{\includegraphics{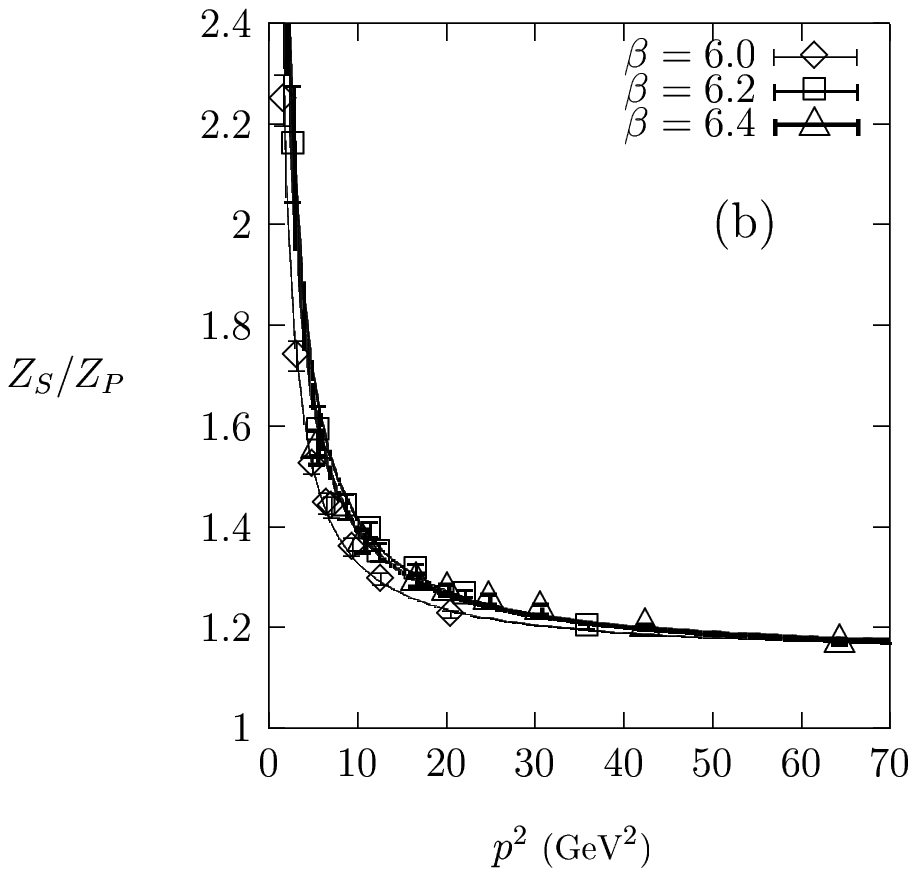}} \par}
\vspace{0.3cm}

\begin{quote}
{\small Figure 1: (a) the data from \cite{giusti2} for \(
Z_{S}^{MOM}/Z_{P}^{MOM} \)
for three values of \( \beta  \), as function of the lattice $a^2 p^2$ and (b)
our fits to these data represented
in {\it physical units}.}{\small \par}
\end{quote}

The comparison of data from the same action at different \( \beta  \) requires
only the ratio of the lattice units \( a \), which are rather well determined.
When we consider the absolute magnitude of the Goldstone term, the uncertainty
becomes larger since the strength (i.e. the coefficient of  $1/(m_q p^2)$ ) is
proportional to \( a^{-3} \). 

Furthermore, a dependence on the action, and therefrom an additional 
cutoff dependence, is to be expected even on the the finite $Z$'s as shown by 
lattice perturbation theory.
In particular, the Wilson term and the clover term induce contributions
to finite $Z$'s of the form $C g^2=C 6.0/\beta$, with a coefficient $C$
dependent on the action. These contributions are quite sizable 
$-$ even for large $\beta$, close to the continuum.
On the other hand, one may expect that
the non-perturbative Goldstone part is independent of the action,
 since it is a long distance effect. Both these expectations are confirmed in
 the present analysis.

\subsubsection{Discretisation errors}

At a given \( \beta  \), one can appreciate discretisation errors, as done in
\cite{giusti1},  by observing
the discrepancies between various quantities which should have been equal, for
instance between various estimates of {\(
(Z_{P}/Z_{S})^{WI} \)}, from various Ward identities (\( WI \)), or else from
the asymptotic value of \( Z_{P}^{MOM}/Z_{S}^{MOM} \). We shall return to their
conclusions at the end of the paper.

However, since one has a series of values for \( \beta  \), it is possible to
do better; by observing the variation
 of a quantity when one increases \( \beta
 \), one can estimate its discretisation error {\it separately}. 
Indeed, the discretisation error will then
correspond to the deviation by powers of $a$ from what is expected close
to the continuum : namely, as we said, one expects a cutoff independent 
Goldstone pole, and a very slow dependence of the perturbative part of  
$Z_P/Z_S$ itself
through $g^2$. 
Note that these expectations amount, on the whole, to saying that $Z_P/Z_S$
should be rather stable with respect to $\beta$, on the  
limited range of available $\beta$ values, and we will refer to this, 
from now on as "stability"; the discretisation errors can then
 be estimated as the deviation from this stability.

Such a study
of
the \( \beta  \) dependence is indeed possible for the Goldstone contribution to
\( Z_{P}^{MOM}/Z_{S}^{MOM}
\)
since one has three \( \beta  \)'s, and to some extent for 
\( Z_{P}^{MOM}/Z_{S}^{MOM} \) itself, although
one must then take into account the
dependence
expected from \( \mathcal{O}(g^{2}) \) corrections, which is very slow, and
one must make sure that the comparisons are performed for the same parameter
values in physical units.

\subsubsection{Extrapolating to $m_q= 0$ }

In the presence of a Goldstone pole, and because of its  \( 1/m_{q} \)
behavior,
there is no
 $m_q\rightarrow 0$  limit at all for ${1/Z_P^{MOM}}$,  and there is the
trivial one \( 0 \) for \( Z_{P}^{MOM} \), or \( Z_{P}^{MOM}/Z_{S}^{MOM} \).
One can define a chiral limit only after subtracting the
Goldstone pole.

On the other hand, if, as in \cite{giusti1,giusti2},
one considers $Z_P^{MOM}$ as it is, without subtraction
of the pole, and if one then makes as usual a linear fit
in $m_q$, {\it the extrapolation to $m_q=0$
is not the chiral limit}. Two questions then arise :\\
1) Is such a linear fit possible, given that the real
behaviour includes a $1/m_q$ term?\\
2)  What is the meaning of the quantity obtained
by this linear extrapolation?

As to question 1), a fit to $1/m_q$ linear in $m_q$  seems possible
for the values of $m_{q}$ reached in standard
numerical simulations, at least for $p^2$ not too small, but is not with
smaller masses, as reached by JLQCD.
As to question 2), we show below that in fact the linear
fit used in \cite{giusti1} gives to a
good approximation \( Z_{P}^{MOM}/Z_{S}^{MOM} \) 
{at an effective mass \( m_{eff}(\beta) \)}
which can be determined by the various masses used in the extrapolation and
which is close to the lowest mass.

\subsection{Results }

{First of all, the extrapolated results} of
\cite{giusti2} can then still be used to observe
the \( 1/p^{2} \) power behavior of the Goldstone pole, and a fit in \( 1/p^{2}
\)
will first allow us to quantify this contribution. Secondly, we shall determine
\( m_{eff}(\beta) \) and this will enable us to observe the typical Goldstone
sensitivity
to the mass: the apparent discretisation errors are in fact due to the hidden
\( m_{eff}(\beta) \) dependence. Finally, we shall be able to confirm the quantitative
estimate of the Goldstone coefficient we have made previously, and to give an
estimate of the true discretisation errors in Wilson data.

\subsubsection{Large power corrections}

To display the power corrections, we perform a fit on the Wilson
 \(
Z_{P}^{MOM}/Z_{S}^{MOM} \) ratio,
given with its statistical errors in \cite{giusti2}, and shown in Fig. 1. Note
that the momentum variable is the true \( p \), not \( \bar{p}={sin(ap)/ a}
\)
as in \cite{giusti1}. The values of the parameters corresponding to the data
are given in Table 1.
\vspace{0.3cm}
{\centering \begin{tabular}{|c|c|c|c|}
\hline
\( \beta  \)&
6.0&
6.2&
6.4\\
\hline
\hline
1/a (GeV)&
2.258\( \pm  \)0.050&
2.993\( \pm  \)0.094&
4.149\( \pm  \)0.161\\
\hline
\hline
\( \kappa _{1} \)&
0.1530&
0.1510&
0.1488\\
\hline
\( \kappa _{2} \)&
0.1540&
0.1515&
0.1492\\
\hline
\( \kappa _{3} \)&
0.1550&
0.1520&
0.1496\\
\hline
\( \kappa _{4} \)&
&
0.1526&
0.1500\\
\hline
\hline
\( \kappa _{crit} \)&
0.15683&
0.15337&
0.15058\\
\hline
\end{tabular}\par}
\vspace{0.3cm}

\begin{quote}
{\small Table 1: the parameters corresponding to the data analysed here, from
\cite{giusti2}.}{\small \par}
\end{quote}
One can clearly see in these data the presence of a pole contribution at small
\( p^{2} \). To quantify it, we fit the points at each \( \beta  \) separately,
to the form:
\begin{eqnarray}
Z_{S}^{MOM}/Z_{P}^{MOM}=a_{S/P}(\beta )+{b(\beta )\over p^{2}}\label{fitSP}
\end{eqnarray}
 with \( b \) in GeV\( ^{2} \) and \( p \) in GeV. The results of this fit
are shown in Fig. 1(b) and in Table 2.

The \( \chi ^{2}/dof \) is rather high, but this is due to the points at high
\( p^{2} \) (this can be verified: cutting out high values of \( p^{2} \)
reduces the \( \chi ^{2}/dof \) substantially), where the discretisation errors
and/or the logarithmic corrections should be the largest. 
As expected from BPT, the $p^2$-independent term 
is remarkably stable with \( \beta  \) (even at 6.0), \( a_{S/P}\simeq 1.14 \).
The coefficient of the power correction is less stable, changing by about 15\%,
but is consistently very large. Hence \( b \) is clearly incompatible with
zero, and gives a very large effect of around \( 30\% \) on the total \(
Z_{S}^{MOM}/Z_{P}^{MOM} \)
at \( 2 \) GeV.

\vspace{0.3cm}
{\centering \begin{tabular}{|c|c|c|c|}
\hline
\( \beta  \)&
6.0&
6.2&
6.4\\
\hline
\hline
\( a_{S/P} \)&
 1.1414\( \pm  \)0.0072&
 1.1266 \( \pm  \) 0.0082&
 1.1364\( \pm  \)0.0049\\
\hline
\( b \)&
 1.8710\( \pm  \)0.063&
2.8996\( \pm  \)0.15&
 2.5844\( \pm  \)0.15\\
\hline
\( \chi ^{2}/dof \)&
1.49&
1.22&
 1.83\\
\hline
\end{tabular}\par}
\vspace{0.3cm}

\begin{quote}
{\small Table 2: the values of the coefficients of Eq. (\ref{fitSP}) fit to
the data of \cite{giusti2}.}{\small \par}
\end{quote}
The consistency of the values of \( b \) at \( \beta =6.2 \) and \( \beta =6.4
\)
shows that this contribution is much beyond the discretisation error on \(
Z_{S}^{MOM}/Z_{P}^{MOM} \).
In fact, these can be estimated to about 2\% from the difference between \(
Z_{S}^{MOM}/Z_{P}^{MOM} \)
at 6.2 and 6.4 at {\( p\simeq 2 \) GeV}.

The difference of \( b \) at \( \beta =6.2 \) and \( \beta =6.4 \) might
be taken as indicating the discretisation artefact \textit{\emph{on the
coefficient
itself}}\emph{.} However, we show in the next section that even this
{difference} is most probably a physical effect, and that the real
discretisation
error on the power correction is still smaller.

\subsubsection{Power corrections are of Goldstone origin}

Having proven the existence of large power corrections, stable with \( \beta
\),
and therefore probably not artefacts of discretisation, we must now prove that
these come from a Goldstone boson. This is in agreement with the dominance of
the divergence of axial current (the pseudoscalar density) at small pion mass,
but in fact we can show that the data itself favours this interpretation.

However, we would 
like first to comment on the dominance of the Goldstone, 
and, for that purpose, to establish the connection with the quantity 
$WIq$ discussed in \cite{giusti1}:
power corrections to \(
Z_{S}^{MOM}/Z_{P}^{MOM} \)
can originate only from the Goldstone boson pole \( \sim 1/m_{q} \), if one
is close enough to the chiral limit.

We start with the Ward identity given in equation (19) of \cite{giusti1}:

\begin{equation}
\label{WIq}
(\frac{Z_{S}}{Z_{P}})^{WI}=\frac{m_{1}\Gamma _{P}(ap;am_{1},am_{1})-m_{2}\Gamma
_{P}(ap;am_{2},am_{2})}{(m_{1}-m_{2})\Gamma _{S}(ap;am_{1},am_{2})}
\end{equation}
 where \( \Gamma _{P} \) and \( \Gamma _{S} \) are the bare vertex functions.
If we assume that \( \Gamma _{P} \) has a Goldstone contribution, whereas \(
\Gamma _{S} \)
doesn't, we get:

\[
\Gamma _{P}=A_{P}(p^{2})+{B_{P}(p^{2})\over m_{q}}+\mathcal{O}(m_{q})\]

\[ \Gamma _{S}=A_S(p^2)(1+\lambda_S(p^2) m_q) \]
 We see that the r.h.s. of Eq. (\ref{WIq}) must then be \( {A_{P}(p^{2})/
A_{S}(p^{2})}+\mathcal{O}(m_{q}) \).
As \( ({Z_{S}}/{Z_{P}})^{WI} \) is a constant, we have:

\[
{A_{P}(p^{2})\over A_{S}(p^{2})}=C+O(m_{q})\]
 with \( C \) a constant, independent of \( p^{2} \).

But in the MOM scheme, the ratio is given by:

\[
\frac{Z_{S}^{MOM}}{Z_{P}^{MOM}}={\Gamma _{P}\over \Gamma _{S}}\]

Hence :

\[ {Z_S^{MOM}\over Z_P^{MOM}}=C+(m_q)^{-1}{B_P(p^2)\over A_S(p^2)}-(m_q)^0{\lambda_S(p^2) B_P(p^2)\over A_S(p^2)}
+
O(m_q)\]

We see that the power corrections are dominated by the Goldstone 
pole $1/m_q$. There seems to be possible additional $O\left((m_q)^0\right)$ power 
corrections, although 
$\lambda_S(p^2)$ is probably small\footnote{This $(m_q)^0$ term
was ommitted in the initial version of
our paper. We thank D. Becirevic 
for having helped us
realise this. That this $\lambda_S$ is indeed
small is implied by the observations of the QCDSF
group for Wilson action, hep-lat/9807044, p.16 ;
for Kogut-Susskind action, it is striking in the
Fig 1 of the JLQCD paper, hep-lat/9901019 ; for
ALPHA action, we thank D. Becirevic for confirming 
that $Z_S$ is incredibly stable with respect to variations
of $m_q$ over a very large range of light masses.
}.
These are themselves roughly
proportional 
to the residue of the Goldstone term, since $\lambda_S(p^2)$ is not
expected to
 have quick variation with $p^2$. At least, they are connected with the
presence
 of the Goldstone pole, and we can say that all the power corrections,
to this
 order $O(m_q^0)$ included, originate in the Goldstone pole. In our
fits, we shall neglect the $\lambda_S(p^2)$ term in Eq. (5), as well as
the smaller $O(m_q)$ terms.

We see also that the the difference :
\begin{equation}
(\frac{Z_{S}}{Z_{P}})^{WI}-\frac{Z_{S}^{MOM}}{Z_{P}^{MOM}}={B_{P}(p^{2})\over
m_{q}A_{S}(p^{2})}-{\lambda_S(p^2) B_P(p^2)\over A_S(p^2)}+O(m_q)
\label{Wiq2}\end{equation}
is entirely due to the Goldstone boson.

\subsubsection{The effective quark mass; Goldstone fit}
{If the $O(m_q)$ corrections are not large, the linear 
extrapolation to $\kappa_{crit}$ which is usually performed amounts 
to making a linear fit in $m_q$
to $1/m_q$, for the 3 or 4 values of}
$m_q$ with \( 2a m_{q}={1/\kappa -1/\kappa _{crit}} \).
The extrapolation of the resulting straight line to \( m_{q}=0 \)
(or to \( \kappa _{crit}) \)
then defines the inverse of an effective mass \( 1/m_{eff}(\beta)\).
The extrapolated \( Z_{S}^{MOM}/Z_{P}^{MOM} \) is thus really calculated, not
at the chiral limit, but at \( m_{eff}(\beta), \) and has the form:
\begin{equation}
\frac{Z_{S}^{MOM}}{Z_{P}^{MOM}}=a_{S/P}(\beta)+\frac{b'}{m_{eff}(\beta)\: p^2},
\end{equation}
in other words:
\begin{eqnarray}
b=\frac{b'}{m_{eff}(\beta)}  & ,
\end{eqnarray}
where $b'$ is a constant, i.e. a number independent of $p$, $m_q$ and $\beta$, which we
call the Goldstone strength.

As long as the Goldstone term is not too large, this result is maintained to
a good approximation when the linear extrapolation is made on the inverse, \(
Z_{P}^{MOM}/Z_{S}^{MOM} \),
which is what is actually done in \cite{giusti2}.

We show in Fig. 2(a) and in Table 3 the result of this extrapolation for the
values of \( \kappa  \) given in Table 1. As we see, \( m_{eff}(\beta) \) is close
to the lowest mass used in the extrapolation.
\bigskip{}

\vspace{0.3cm}
{\centering \begin{tabular}{|c|c|c|c|}
\hline
\( \beta  \)&
6.0&
6.2&
6.4\\
\hline
\hline
\( m_{eff}(\beta) \) (GeV)&
 0.0591&
0.0400&
0.0434\\
\hline
\hline
\( b' \) (GeV\( ^{3} \))&
0.1106\( \pm 0.004 \)&
0.116\( \pm 0.006 \)&
0.112\( \pm 0.006 \)\\
\hline
\end{tabular}\par}
\vspace{0.3cm}
\begin{quote}
{\small Table 3: the values of the effective mass defined by the linear
extrapolation,
and the resulting values of the coefficient of the Goldstone term.}{\small
\par}
\end{quote}
\vspace{0.3cm}
{\par\centering \resizebox*{0.49\textwidth}{!}{\includegraphics{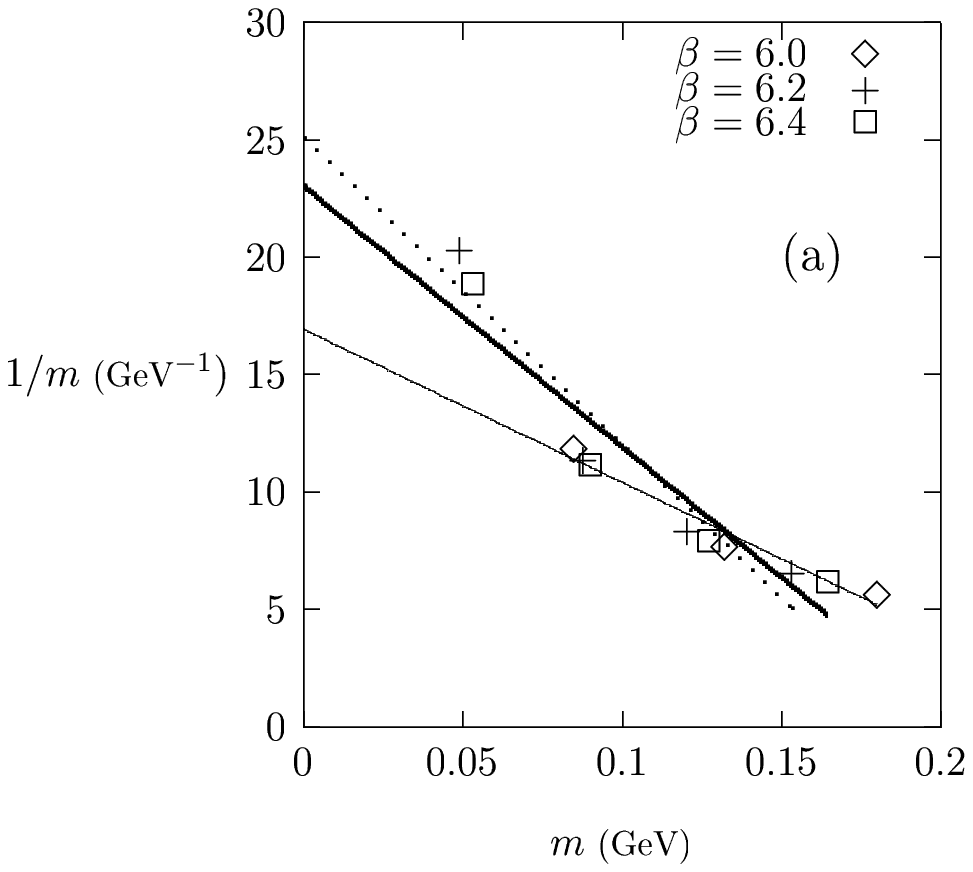}}
\resizebox*{0.49\textwidth}{!}{\includegraphics{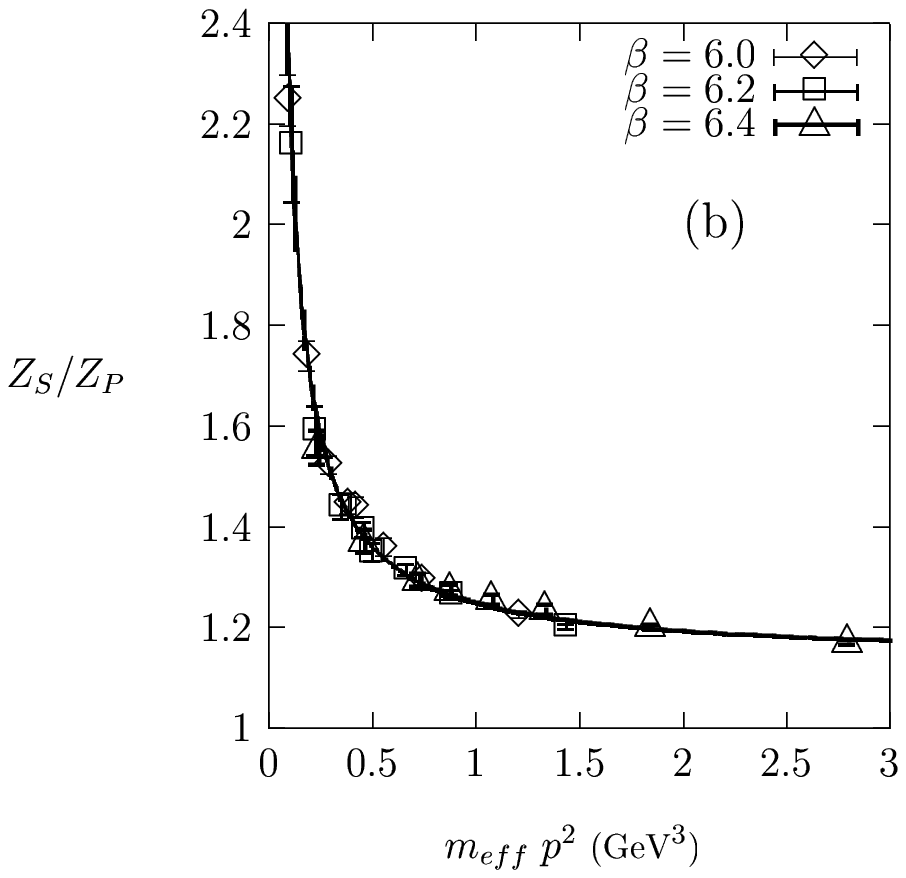}} \par}
\vspace{0.3cm}

\begin{quote}
{\small Figure 2: (a) 
The three straight lines extrapolate the values of $ 1/m_{q} $
at each $\beta$. Their intersection with the $m=0$ ordinate defines
$ 1/m_{eff}(\beta) $  and (b) the result of a joint fit to all data, after
the $ 1/m_{eff}(\beta) $ dependence of the Goldstone term has been taken into
account.\par}\bigskip{}

\end{quote}
\vspace{0.3cm}

We can now deduce \( b' \) from the fit to \( b \) calculated previously:
\( b'=bm_{eff}(\beta) \). The striking result, shown in Table 3, is that the rather
different values of \( b \) obtained previously in Table 2 correspond to very
good approximation to the same \( b' \), i.e., the {Goldstone strengths}
extracted from the data are almost the \textit{same}, and the difference of
the \( b \)'s is mainly due to the different values of \( m_{eff}(\beta) \) for each
\( \beta  \), which are due to the choice of values of \( \kappa  \).

This effect is also evident
in Fig. 5 of
\cite{giusti2} and in our Fig. 1(b)
, where \( Z_{P}^{MOM}/Z_{S}^{MOM} \) at \( \beta =6.0 \), with \( p \) in
physical units, deviates significantly from its value at the higher \( \beta
\)'s,
at small momenta. Such a deviation cannot be explained by discretisation
effects,
which are not confined at small momenta. The natural explanation is that the
effective mass is notably higher than at higher \( \beta  \), and that the
apparent deviation of \( Z_{P}^{MOM}/Z_{S}^{MOM} \) at \( \beta =6.0 \) is
\textit{\emph{almost
entirely due to the quark mass dependence of the Goldstone effect}}, and
disappears
when results are compared not only at same physical \( p \), but also at same
physical quark mass. The typical quark mass dependence of the Goldstone had
been hidden by the extrapolation procedure, but has reappeared as a completely
spurious discretisation effect. The true Goldstone origin of this fictitious
discretisation effect is revealed by its {\(
1/(p^{2}m_q) \) behaviour}.

We are now in a position to perform a joint fit to the data at the three values
of \( \beta  \), with the variable \( 1/(m_{eff}(\beta)\: p^{2}) \) instead of \(
1/p^{2} \).
We find, with a \( \chi ^{2}/dof=1.39 \):
\begin{eqnarray}
Z_{S}^{MOM}/Z_{P}^{MOM}=a_{S/P}(\beta)+{(0.112\pm 0.025)\;
\textrm{GeV}^{3}\over m_{eff}(\beta)\: p^{2}} & \label{fit1}
\end{eqnarray}

The dependence of $a_{S/P}$ on $\beta$ is very weak: we find 
$1/a_{S/P}(\beta)=0.88\pm 0.05$, $0.88\pm 0.04$ and $0.88\pm 0.03$ respectively
at $\beta=6.0$, $6.2$ and $6.4$.

The Goldstone contribution is stable, compatible at the 3 \( \beta  \)'s
within statistical errors, and very large when \( p\simeq 2 \) GeV and
\( m_{q}\simeq 50 \) MeV. Note that these results, shown in Fig. 2(b), are
in direct contradiction with the conclusions of \cite{giusti1}.

The mildness of cutoff dependence is manifest in the possibility of making such
a good common fit to the data for the three different \( \beta  \)'s. This
possibility also gives strong support to the Goldstone interpretation, since
it would not be possible without accounting for the \( 1/m_{q} \) dependence.

\subsubsection{\protect\( a^{2}p^{2}\protect \) discretisation errors}

 From the fit (\ref{fit1}), we can deduce
the asymptotic value of 
\(Z_{P}^{MOM}/Z_{S}^{MOM}=1/a_{S/P} \approx 0.88 \),
close to the BPT result. This should be equal to the value given by the Ward
identity
(\ref{WIq}), from which however one gets a lower result \( 0.79-0.80 \)
\cite{giusti1}.


In fact, 
there seems 
to be a \( p \)-dependent effect, which is signaled
by the fact that the strength of the Goldstone term
becomes slightly lower and that
the \( \chi ^{2} \) improves when cutting off the large \( a^{2}p^{2} \)
points.

Also, correspondingly,
the right-hand side of Eq. (\ref{WIq}), which
equates to \( Z_{S}^{MOM}/Z_{P}^{MOM} \) minus the Goldstone, is not perfectly
constant
as it should, although it is {much more so}
 than \( Z_{P}^{MOM}/Z_{S}^{MOM} \)
itself,
illustrating the Goldstone interpretation (see Fig. 3 of \cite{giusti1}). It
is also somewhat different from \( 1/a_{S/P} \), although the comparison
between
a constant and a varying quantity is difficult.

The residual cutoff dependence (leaving aside the \( \mathcal{O}(g^{2}) \)
effect) can be fitted by a small \textit{negative} \( a^{2}p^{2} \) term in
\( Z_{S}^{MOM}/Z_{P}^{MOM} \). We can then obtain very good fits at the three
\( \beta  \)'s with \textit{a universal Goldstone strength \( b'=0.098 \)}
\textit{\emph{GeV}}\( ^{3} \) and a constant term \( a_{S/P} \) which decreases
with \( \beta  \) just as expected from BPT, and we obtain our final result:
\begin{eqnarray}
Z_{S}^{MOM}/Z_{P}^{MOM}=a_{S/P}(\beta )+{(0.098\pm 0.004)GeV^{3}\over m_{eff}(\beta)\:
p^{2}}-(0.013\pm 0.003)a^{2}p^{2} &
\end{eqnarray}
 with \( \chi ^{2}/dof=0.45 \) and \( dof=19 \). The corresponding values
of \( a_{S/P} \), together with the expectations from BPT and the Ward
identities,
are shown in Table 3. From this table, it is visible that, once more, after due
subtraction of the essentially non perturbative Goldstone pole effect, one has
a result close to BPT. Note however that the BPT estimate quoted here is the
ratio of the one-loop
BPT estimates of $Z_P$ and $Z_S$ at $a p=1$; one would obtain a somewhat
different result, and one exactly scale independent, by applying one loop BPT
directly to the ratio.

\vspace{0.3cm}
{\centering \begin{tabular}{|c|c|c|c|}
\hline
\( \beta  \)&
6.0&
6.2&
6.4\\
\hline
\hline
\( 1/a_{S/P} \) (our fit)&
0.835\( \pm  \)0.010&
0.845\( \pm 0.009 \)&
0.845\( \pm 0.007 \)\\
\hline
\( 1/a_{S/P} \) (BPT) \cite{giusti2}&
0.83&
0.84&
0.85\\
\hline
\( 1/a_{S/P} \) (WIq) \cite{giusti1}&
 -&
\( 0.79\pm 0.02 \) &
\( 0.80\pm 0.02 \)\\
\hline
\end{tabular}\par}
\vspace{0.3cm}

\begin{quote}
{\small Table 3: our determination of \( 1/a_{S/P} \) compared with other
determinations from \cite{giusti2},\cite{giusti1}.}{\small \par}\bigskip{}
\end{quote}

We show in Fig. 3 that the Goldstone contribution dominates
the discretisation artefact described by our $a^2 p^2$ term, up to the highest
considered \( a^{2}p^{2} \) for the two higher values
of \( \beta  \).

\vspace{0.3cm}
{\par\centering \resizebox*{0.6\textwidth}{!}{\includegraphics{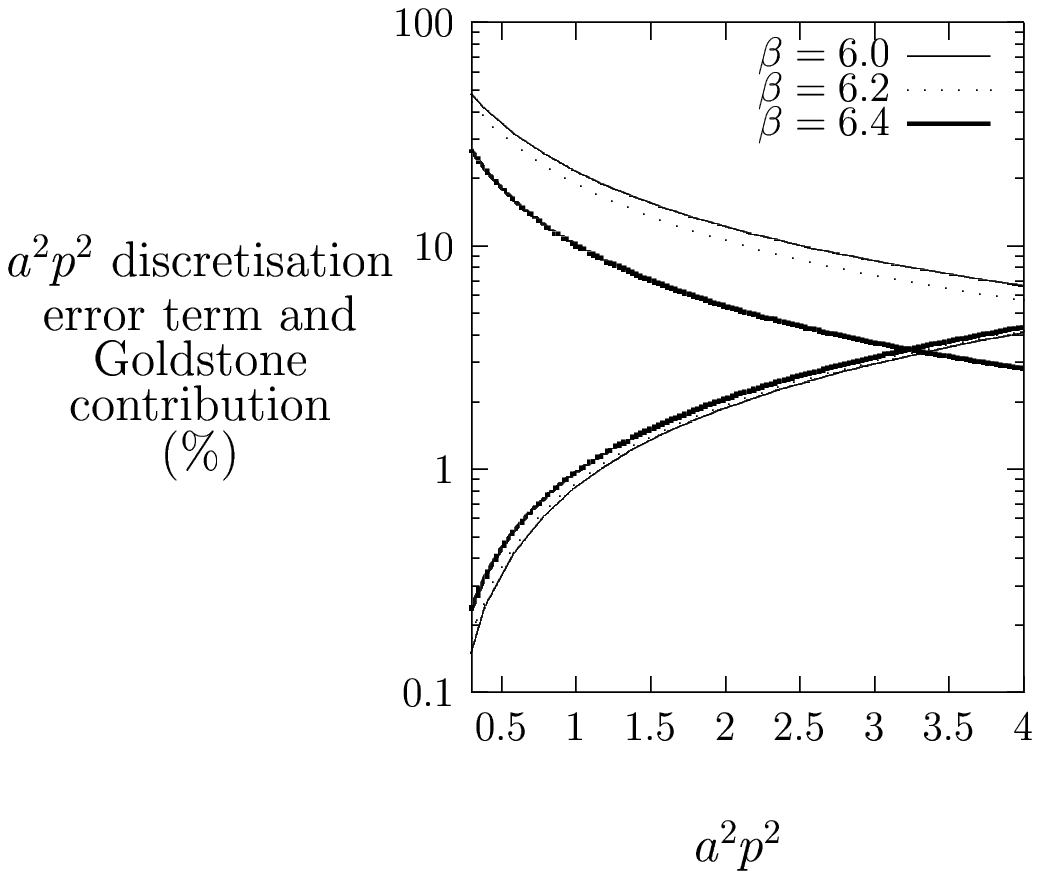}}
\par}
\vspace{0.3cm}

\begin{quote}

Figure 3: the contribution of the $a^2 p^2$ term (rising curves)
and of the Goldstone contribution (decreasing curves) relative to the total
${Z_S\over Z_P}$.{\small \par}\bigskip{}
\end{quote}
\vspace{0.3cm}
This fit, as shown in Table 3, also gives an estimate of the asymptotic
contribution
to the Ward Identity, \( 1/a_{S/P} \), lower and closer to (\ref{WIq}).
We also give our estimate of the relative sizes of the three terms at
$a^2p^2=1$ in Table 4.

\vspace{0.3cm}
{\centering \begin{tabular}{|c|c|c|c|}
\hline
\( \beta  \)&
6.0&
6.2&
6.4\\
\hline
\hline
Goldstone term&
 21.5\% &
19.0\%&
10.1\%\\
\hline
\hline
$a^2p^2$ discretisation errors&
-0.83\%&
-0.87\%&
-0.96\%\\
\hline
\end{tabular}\par}
\vspace{0.3cm}
\begin{quote}
{\small Table 4: the relative values of the Goldstone boson and of the
$a^2p^2$ discretisation errors at $a^2p^2=1$.
}
\end{quote}

We
must of course keep in mind that the procedure used here is crude, and that
a proper analysis of the data must take the Goldstone contribution into account
for each \( \kappa  \), before extrapolating to the chiral limit. Hence the
agreement with BPT and the difference with the WI determination may reflect
the crudeness of our method, due to the unavailability of better data. It is
also possible to get {fits still closer} to \(
1/a_{S/P}=0.8 \) by allowing for $a^4 p^4$ terms, at the cost of some variation
of $b'$ with $\beta$. On the other hand it is clear that a similar analysis of
discretisation artefacts of the type $a^n p^n$ could be usefully applied to
$WIq$, which does not appear to be perfectly constant. At any rate, we observe
that even with our latter fits with $a^4 p^4$ terms, the main conclusion is
still the same: the Goldstone strength is large, close to our previous
estimate, and well above any estimate of discretisation errors over a large
range of momenta above $2~{\textrm {GeV}}$.

\subsubsection{Comparison with Goldstone residue extracted from QCDSF improved
data}

We can compare the Goldstone residue with our result of Eq. (\ref{notreZP2}),
extracted from the QCDSF data. Since we do not know {the
value} of \( Z_{S} \)
from the ALPHA action, the simplest thing to do, disregarding slow logarithmic
evolutions, is to compare the magnitudes of the ratio of the power correction
to
the perturbative term at \( p \) and \( m_q \) similar in physical units, or
of the Goldstone strength $b'$ to the perturbative term in the same unit
GeV\(^{3} \).
We immediately notice that the latter ratio is in perfect agreement:  0.82
GeV\(^{3} \)
for the improved action, \( 0.818\pm 0.003 \) GeV\( ^{3} \) for the Wilson
action.

Hence there is \textit{\emph{full compatibility between the various
determinations
of the Goldstone strength, both at various \( \beta  \)'s and for various
actions,
converging towards very large values, some \( 30\% \) of the total around \( 2
{}~{\textrm {GeV}} \)
and for a mass around \( 50 \)}} MeV.

\section{Discretisation errors; an overall discussion}

We now come to the paper {\cite{giusti1} of Giusti and Vladikas (G\&V)},
and {to a discussion of the origin of its conclusions, opposite to
ours.} Admittedly, many of the procedures used there are common in the
literature,
{but they turn out to be} inappropriate for
the present discussion of the Goldstone contribution. Hence we think that
beyond answering the {criticisms} of \cite{giusti1}, commenting
upon them is of general interest.

Although they {intended} to discuss specifically our estimate of the
magnitude of the Goldstone contribution, the core of the argument of
{G\&V} is the comparison of various
determinations
of \( (Z_{P}/Z_{S})^{WI} \).
{The spread of the values directly extracted
from Ward identities, and the deviation with the values obtained for   \(
Z_{P}^{MOM}/Z_{S}^{MOM} \),} are supposed to be a measure
of the discretisation error on \( Z_{P}^{MOM}/Z_{S}^{MOM} \)
{itself. This way of estimating the errors
is one of our main disagreements,} for reasons expressed below in points 2 and
3. {The other main difference is }
that {G\&V} do not take into account the strong scale
dependence of \( Z_{P}^{MOM}/Z_{S}^{MOM} \), as explained in points 1 and 4.

\begin{enumerate}
\item
Let us first emphasize that, even admitting the
$10-15~\%$ estimate of discretisation errors made by \cite{giusti1}, these
errors cannot dominate
the $30~\%$ estimate of the Goldstone pole that we have given at $2~{\textrm
{GeV}}$. The reason why {G\&V have}  missed this point
is clear. The argument, illustrated in their Fig. 4 and Table 1, relies on a
point with large momentum for
each \( \beta  \). The numbers of ref. \cite{giusti1} are given at sin\(
^{2}(ap)=0.8 \),
i.e. at \( p=3.3 \) GeV \( (\beta =6.2) \) or \( p=4.6 \) GeV \( (\beta =6.4)
\).
This choice of a large physical momentum is not {appropriate}
when the manifest goal
is to discuss the Goldstone pole overall strength. Indeed, if there is a
Goldstone pole, \( Z_{P}^{MOM}/Z_{S}^{MOM} \)
strongly depends on \( p^{2} \), and its difference with the asymptotic value
decreases rapidly with increasing \( p^{2} \), rendering difficult or
eventually impossible the determination of the power correction. Had
{G\&V} taken \( p=2\) GeV,
they would have had to quote a central value for \( Z_{P}^{MOM}/Z_{S}^{MOM} \)
around
0.6 or less ({as can be seen from} their Fig.~3), much below
\( WIq=0.79 \) and also below \( WIh=0.68 \) $-$ making manifest the large
magnitude of the Goldstone {$-$}. In the introduction to their new
version of the paper \cite{giusti1}, they state however that \(
Z_{P}^{MOM}/Z_{S}^{MOM} \) is compatible with the $WI$'s even at $2~{\textrm
{GeV}}$ within discretisation errors; this, from their own numbers, amounts to
admitting still larger discretisation errors of 
the order of $30~\%$ or more at
$\beta=6.2$ and $a^2 p^2=0.45$ (difference between $0.79$ and $0.6$).
One the other hand, the only known way to explain the data
with a reasonable error estimate is through the Goldstone interpretation.
\item
Furthermore, let us emphasize that their estimated
discretisation errors contradict the evolution of data with
$\beta$ as far as \( Z_{P}^{MOM}/Z_{S}^{MOM} \) is concerned.
Indeed in
\cite{giusti2}, the statistical errors on \( Z_{P}^{MOM}/Z_{S}^{MOM} \) are
small, and the discretisation errors seem also small, since the values of
\( Z_{P}^{MOM}/Z_{S}^{MOM} \) \textit{\emph{taken at the same physical
momenta}}
differ only by a few percent between \( \beta =6.2 \) and \( \beta =6.4 \)
(see Fig. 1(a)).

 A careful reading of the text reveals that the
much larger error introduced in \cite{giusti1}  \textit{\emph{has actually
nothing
to do with the error on \( Z_{P}^{MOM}/Z_{S}^{MOM} \) itself}}, but really
concerns the estimated error made on the indirect estimate of the Ward identity
result through the asymptotic value of \( Z_{P}^{MOM}/Z_{S}^{MOM} \). This
error was already given
in \cite{giusti2}, and in fact, in  \cite{giusti2}, the {\it same} numbers were
quoted as ``RGI'' (i.e., estimate of the asymptotic, renormalisation group
invariant quantity). Indeed, in \cite{giusti2}, the lack of the expected
plateau
was interpreted as an error of {10 to 15\%} on
$(Z_{P}/Z_{S})^{RGI}$. In \cite{giusti1}, the same number is re-expressed
arbitrarily
as an error  on the value of \( Z_{P}^{MOM}/Z_{S}^{MOM} \) at the lower end of
the range, \( a^{2}\mu ^{2}=0.8 \), though it is not an actual error on \(
Z_{P}^{MOM}/Z_{S}^{MOM} \).

In our opinion, the lack of plateau signals power corrections and the need to
subtract them.
{The procedure of \cite{giusti1}, which amounts to including them
automatically into the errors on \( Z_{P}^{MOM}/Z_{S}^{MOM} \), makes it of
course impossible to discuss the power corrections.}
\item
{Moreover, G\&V} further substantiate their estimate of the
discretisation
effect by observing a $10-15\%$ discrepancy between determinations from two
Ward
identities, called \emph{WIq} and \emph{WIh}, at \( \beta =6.2 \). This
procedure
has the advantage that the Ward identities are scale independent. It is also
a natural approach, if one is working at only one \( \beta  \), to look for
the difference between quantities which should be equal. However, in this
approach,
one does not know which is the best estimate, or whether both equally fail: the
same discretisation error is attributed to both, and to any other quantity,
such as \( (Z_{P}/Z_{S})^{RI/MOM} \), which may be over-pessimistic.

As already emphasized, a better approach is to examine the 
variation of the
specific  quantity  which one wants to study, when one increases
\( \beta  \). Admittedly, small variations can present 
from the variation of $\alpha_S$, 
but they should be $O(g^2)$ in BPT and vary slowly with $\beta$, 
and hence the quantity
should be very stable when $\beta$ changes. 
If it is stable up to logarithms, 
this particular quantity has probably a small
discretisation error. It seems that
the \emph{WIq} determination changes very slightly  between
6.2 and 6.4, from 0.79 to 0.8. In fact, the values are compatible within
statistical
errors, and the small increase is expected from BPT. This is not so for
\emph{WIh},
which shows a strong variation from 0.68 to 0.73. In fact, \emph{WIh}
corresponds
to \( Z_{A} \) times the ratio \( \rho /m_{q} \) of the axial to the subtracted
mass, and the latter ratio is known to exhibit rather large variations. The
natural conclusion would be then that \emph{WIq} deserves more trust than
\emph{WIh}.
The same can be said probably of \( Z_{P}^{MOM}/Z_{S}^{MOM} \), which is
remarkably stable, with a small variation in agreement with BPT.
.  Thus the large discretisation error should be probably
attributed to \emph{WIh} only, not to the three quantities at the same time. It
is then rewarding that \emph{WIq} and \( Z_{P}^{MOM}/Z_{S}^{MOM} \) give
compatible results for the estimate of $WI$, after due subtraction of the
Goldstone pole, as we have shown above.
\item
We note that in their Fig. 4 and Table 1,
{G\&V} consider the spread of
values of various determinations of \( Z_{P}/Z_{S} \) with increasing \( \beta
\), including \( Z_{P}^{MOM}/Z_{S}^{MOM} \),
at the same \( a^{2}\mu ^{2}=0.8 \), and not at at the same physical \( p \),
as one should do {when discussing} the
error on a {scale-dependent}
quantity. {G\&V} are in fact not comparing the same quantity
at two different
\( \beta  \)'s, but \textit{\emph{two different quantities}}: the values of \(
Z_{P}^{MOM}/Z_{S}^{MOM} \)
respectively at \( p=3.3 \) GeV and \( p=4.6 \) GeV.

The natural explanation of the increase of \( Z_{P}^{MOM}/Z_{S}^{MOM} \) with
\( \beta \) in their figure and table is the decrease of the power correction
with increasing \( p \), which
is a physical effect, not the discretisation errors, except at very large \( p
\). If we duly compare at identical \( p \), we see once again
that \( Z_{P}^{MOM}/Z_{S}^{MOM} \) is very stable with \( \beta  \), and that
the discrepancy with \( WIq \) does not decrease, unlike suggested
by the Fig.~4 of \cite{giusti1}: it is a physical effect, the sign of Goldstone
contribution as shown above.
\end{enumerate}

\section{Conclusion}

The quantity \( Z_{P}^{MOM}/Z_{S}^{MOM} \) appears to be the best indicator
of the Goldstone pole. Contrarily to \cite{giusti1}, we find a large Goldstone
contribution to the Wilson data, of the same magnitude as found previously with
data for \( Z_{P}^{MOM} \) from the QCDSF improved action. The results are
consistent
for 3 values of \( \beta  \) and for momenta ranging from about \( 1 \) GeV
to \( 8 \) GeV. Of course, the determination of the Goldstone strength comes
mainly from moderate momenta, where the contribution is the largest. The
discretisation uncertainty, as estimated from the variation of \(
Z_{P}^{MOM}/Z_{S}^{MOM} \) with $\beta$, appears in
fact to be rather small, and the Goldstone contribution at \( p=2-4 \) GeV  is
far above it. Even
admitting the larger discretisation error advocated by \cite{giusti1}, which is
not relevant in our opinion, our claimed Goldstone contribution at $2~{\textrm
{GeV}}$ is so large, as already found previously, that it is clearly
dominating. Evidently, it is smaller at the higher momenta considered by
{G\&V} in their Fig. 4 and Table 1, but this is as it should
be: it must be $\propto 1/p^2$ !

This large Goldstone contribution explains in a natural manner the discrepancy
of the MOM $Z_P$ with the ALPHA group determination of $Z_P$ at large distance
(around $30\%$ at $2~{\textrm {GeV}}$). It also explains for the most part the
absence of
plateau in \( Z_{P}^{MOM}/Z_{S}^{MOM} \), even at the highest acceptable
momenta. True, we find some contribution from $a^2 p^2$ artefacts, but
certainly not a dominant one.
Given this absence of a plateau, one should not insist on extracting an
estimate of \( (Z_{P}/Z_{S})^{WI} \)
directly from \( Z_{P}^{MOM}/Z_{S}^{MOM} \), even if
one assumes large errors. The only way to proceed, which we have illustrated
here, is to \textit{\emph{subtract}} the Goldstone contribution. The result of
the subtraction is, once more, a number close to the BPT expectation, which is
quite encouraging. Another formulation
of this is to use Eq. (\ref{WIq}), which automatically subtracts the Goldstone,
and which is found to be rather stable with \( \beta  \).

Of course, {some
slight changes in the conclusions must be expected} from a more
thorough analysis of the complete data, where it may be
possible, in particular,
to explain the small discrepancy between \( WIq \) and \( 1/a_{S/P} \) given
in Table 3.

It remains to explain the apparently different conclusion from \cite{ROME},
at \( \beta =6.2 \), which find smaller power corrections; this may be
related to \textit{\emph{off-shell}} improvement.

The interesting and intriguing physical question is now to find the reason why
the Goldstone residue is so large in \( Z_{P}^{MOM} \) or \(
Z_{P}^{MOM}/Z_{S}^{MOM} \).
This is connected with the behavior of the pion wave function (BS amplitude)
at short distance. The apparent contradiction with the standard OPE is
puzzling.
 A naïve interpretation of our results\footnote{
Note that, contrarily to what was written in the
first version of \cite{giusti1}, we have never
ourselves proposed
such an interpretation. Note however that in the mechanism of spontaneous color
symmetry breaking proposed recently by
C. Wetterich  \cite{Wetterich}, the propagator could get
large fluctuations in the octet and this could be
connected with our
observation (private communication).}
would be to claim that the quark condensate is 10 times the standard value,
but this is certainly not probable, and the ultimate physical reason of this
disagreement must surely be more subtle.

\section*{Acknowledgments: }

We are very much indebted to Damir Becirevic for numerous discussions 
and suggestions,
and very useful criticism. We thank also Guido Martinelli,  as
well as the Quadrics group, especially J.-P. Leroy, and also the authors of
\cite{giusti1} for important discussions; 
we also thank S. Capitani, V. Lubicz and R.
Sommer for previous comments on the question. This research is supported in
part by the CNRS-CGRI cooperation agreement 99-11.

\section*{Note added in proof: }
While we were writing this letter, C. Dawson \cite{Dawson:2001kh}
and Y. Zhestkov \cite{Zhestkov:2001bs}
stressed again the necessity
of subtracting the Goldstone pole to obtain
a chiral limit.

\end{document}